\documentclass[aps,preprint,preprintnumbers,nofootinbib,showpacs]{revtex4}
\usepackage{graphicx,color,amsmath}

\newcommand{\newc}{\newcommand}
\newc{\be}{\begin{equation}}
\newc{\ee}{\end{equation}}
\newc{\beq}{\begin{eqnarray}}
\newc{\eeq}{\end{eqnarray}}

\begin{document}

\title{Cosmological birefringence induced by neutrino current}
\author{C. Q. Geng$^{a,b}$, S. H. Ho$^{a,b}$ and J. N. Ng$^b$}
\address{$^a$Department of Physics, National Tsing Hua University, Hsinchu, Taiwan 300\\
$^b$Theory group, TRIUMF, 4004 Wesbrook Mall, Vancouver, B.C. V6T 2A3, Canada}
%\author{}
%\address{Theory group, TRIUMF, 4004 Wesbrook Mall, Vancouver, B.C. V6T 2A3, Canada}

%\shortauthor{C. Q. Geng, S. H. Ho and J. N. Ng}

%
\begin{abstract}
We review our recent work \cite{Geng:2007va} on the 
cosmological birefringence.
We propose a new type of effective interactions in terms of the
$CPT$-even dimension-six Chern-Simons-like term to generate
the cosmological birefringence.
We use the
neutrino number asymmetry to induce  a non-zero
rotation polarization angle in the  data of the
cosmic microwave background radiation polarization.
%\\\\PACS Nos.: 
\end{abstract}
\maketitle

\section{Introduction}

The polarization maps of the cosmic microwave background (CMB)
have been  important tools for probing the epoch of the last
scattering directly. As we know, the polarization of the CMB can only
be generated by Thomson scattering at the last scattering surface
and therefore linearly polarized \cite{r-9,r-10}. When a
linearly polarized light travels through the Universe to Earth,
the angle of the polarization might be rotated by some localized
magnetized plasma of charged particles such as ions and electrons,
this is so-called Faraday effect. However, the rotated angle of
the polarization plane by this Faraday effect is proportional to the
square of the photon wavelength and thus it can be extracted.

On the other hand,
%about ten years ago 
in 1997 Nodland and Ralston \cite{r-1} claimed
that they found an additional rotation of synchrotron radiation
from the distant radio galaxies and quasars, which is
wavelength-independent and thus different
from  Faraday rotation, referred as the cosmological birefringence.
Unfortunately, it has been shown that there is no statistically significant signal present \cite{r-4,Comments}.
Nevertheless, this provides a new way to search for new physics in
cosmology.
Recently, Ni \cite{WTNi-pol} has pointed out that the change of the rotation angle of the polarization can be constrained at the
level of $10^{-1}$ by
 the data of
the Wilkinson Microwave Anisotropy Probe (WMAP) \cite{WMAP}
due to the correlation between the polarization and temperature.
 Feng $et\ al$ \cite{r-2} have used the combined data of the WMAP and
 the 2003 flight of BOOMERANG (B03) \cite{B03}
 for the CMB polarization to further constrain
the rotation angle  and concluded that
a nonzero angle is mildly favored. 
%Note added: After the completion of this work, there was an
%interesting paper by
Recently, Cabella, Natoli and Silk \cite{Silk}, 
%which
%gives a constraint
 have applied a wavelet based estimator on the WIMAP3 TB and EB data
 to constrain the cosmological birefringence.
 They derive a limit of $\Delta\alpha=-2.5\pm3.0$ deg, which is slightly
 tighter than that in Ref. \cite{r-2}.
For a more general dynamical scalar, this rotation angle is
more constrained \cite{Liu}.
If such rotation angle does exist, it clearly
 indicates an
anisotropy of our Universe.
%It is known that t
This phenomenon can be also used to test the Einstein equivalence principle
% as was first pointed out by Ni
 \cite{WTNi,WTNi-r}. 
 
 In this talk, we will review our recent work on the
 cosmological birefringence \cite{Geng:2007va}. 
 In Ref. \cite{Geng:2007va}, we propose a new type of effective interactions in terms of the $CPT$-even dimension-six Chern-Simons-like term to generate the cosmological birefringence.
In particular, we use the
neutrino number asymmetry to induce  a non-zero
rotation polarization angle in the  data of the
cosmic microwave background radiation polarization.
%\\ 

\section{Dimension-six Chern-Simons-like Lagrangian with Neutrino Current} 
%Theoretical explanation}

One of interesting theoretical origins for the birefringence was  developed by Carroll, Field and Jackiw (CFJ)
 % $et\ al$ \cite{r-4,r-3}.
 \cite{r-3}. They
modified the Maxwell Lagrangian by adding a Chern-Simons term \cite{r-3}:
\begin{eqnarray} \label{lagrangian}
\cal{L}&=&\cal{L}_{EM}+\cal{L}_{CS}
\nonumber \\
          &=&-\frac{1}{4}\sqrt{g}\emph{F}_{\mu\nu}\emph{F}^{\mu\nu}-\frac{1}{2}\sqrt{g}\emph{p}_{\mu}\emph{A}_{\nu}\emph{\~{F}}^{\mu\nu}\,,
\end{eqnarray}
where
$\emph{F}_{\mu\nu}=\partial_{\mu}A_{\nu}-\partial_{\nu}A_{\mu}$ is the electromagnetic tensor,
$\emph{\~{F}}^{\mu\nu}\equiv\frac{1}{2}\epsilon^{\mu\nu\rho\sigma}\emph{F}_{\rho\sigma}$
is the dual electromagnetic tensor, g is defined by
g=-det($g_{\mu \nu}$), and
 $p_{\nu}$ is a
four-vector.
 Here,  to describe a
flat, homogeneous and isotropic universe,
we use the Robertson-Walker metric
\begin{equation}
ds^2=-dt^2 +R^2(t)\;d\textbf{x}^2\,,
\end{equation}
 where R is the scale factor;
and the totally anti-symmetric tensor Livi-Civita tensor
$\epsilon^{\mu\nu\rho\sigma}=g^{-1/2}e^{\mu\nu\rho\sigma}$ with
the normalization of $e^{0123}=+1$.

 In the literature \cite{r-4,r-3,Carr1,Carr2,r-5,XZhang0611}, $p_{\mu}$ has been taken as a constant  vector or the gradient of a scalar.
In this paper, we study the possibility that the four-vector
$\emph{p}_{\mu}$ is related to a neutrino current
\begin{eqnarray}
\emph{p}_{\mu}&=&\frac{\beta}{M^2}\emph{j}_{\mu}\,
\end{eqnarray}
with the four-current
\begin{eqnarray}
\emph{j}_{\mu}&=&\bar{\nu}\gamma_{\mu}\nu\;\equiv\;(j^{0}_{\nu}, \vec{\emph{j}_{\nu}})\,,
\label{nu-current}
\end{eqnarray}
 where $\beta$ is the coupling constant of
order unity and M is an undetermined new physics mass scale.
Note that $\vec{\emph{j}_{\nu}}$
is the neutrino flux density and $j^{0}_{\nu}$ is the number density difference between neutrinos and anti-neutrinos, given by
%The neutrino current is $\emph{j}_{\mu}=\bar{\nu}\gamma_{\mu}\nu =
%(\Delta n_{\nu} , \vec{\emph{j}_{\nu}})$, here $\Delta
%n_{\nu}=n_{\nu}-n_{\bar{\nu}}$ is the number density difference
%between neutrinos and anti-neutrinos, and $\vec{\emph{j}_{\nu}}$
%is the neutrino flux density.
\begin{eqnarray}
j^{0}_{\nu}&=&\Delta n_{\nu}\;\equiv\; n_{\nu}-n_{\bar{\nu}}\,,
\label{Delta-n}
\end{eqnarray}
where $n_{\nu(\bar{\nu})}$ represents the neutrino (anti-neutrino)
number density. It should be noted that if $\Delta n_{\nu}$ in Eq. (\ref{Delta-n}) is nonzero, the cosmological
birefringence occurs even in the standard model (SM) of particle interactions \cite{pal}. However, the effect is expected to be vanishingly small \cite{pal}.
In the following discussion, we will ignore this standard model effect.

As we are working on the
usual Robertson-Walker metric,
the particle's phase space distribution function
is spatially homogeneous and isotropic, i.e. $\textit{f}
(p^{\mu},x^{\mu})$ reduces to $\textit{f} (\mid \vec{p}\mid,t)$
or  $\textit{f} (E,t)$  \cite{r-7}. In other words,
the relativistic neutrino background in our Universe is assumed
to be
homogeneous and isotropic like the CMB
%cosmic microwave background
radiation, which implies that
the number density for neutrinos is only a function of
red-shift z, i.e. the cosmic time. As a result, we conclude that the
neutrino current in Eq. (\ref{nu-current})
to a co-moving observer has the form
\begin{eqnarray} \label{current}
 \emph{j}_{\mu}&=&
 %\bar{\nu}\gamma_{\mu}\nu =
 %\bigg{(}\Delta n_{\nu}\big{(}z(t)\big{)}, - D \vec{\nabla}(\Delta
% n_{\nu}\big{(}z(t)\big{)}\bigg{)} \\ \nonumber
%&=&
\bigg{(}\Delta n_{\nu}\big{(}z(t)\big{)}, \vec{0}\bigg{)}\,.
\end{eqnarray}
Note that $\vec{j}=- D \vec{\nabla}\big[\Delta n_{\nu}\big{(}z(t)\big{)}\big]$,
 where D is diffusivity \cite{r-8} and $\vec{\nabla}$ is the
usual differential operators in Cartesian three-space. Here, we
have constrained ourselves to consider only the relativistic
neutrinos (for homogeneous and isotropic).

As pointed out by Carroll $et\ al$ \cite{r-3}, in order to preserve the gauge invariance we must require that the variation of $\cal{L}_{CS}$,
given by
\beq
\cal{L}_{CS}&=&
-\frac{1}{2}\sqrt{g}{\beta\over M^{2}}j_{\mu}\emph{A}_{\nu}\emph{\~{F}}^{\mu\nu}\,,
\label{Lcs}
\eeq
vanishes under the gauge transformation of $\Delta A=\partial_{\nu}\chi $
for an arbitrary $\chi$.
To check the  gauge invariance in our Lagrangian of Eq. (1),
we write
\begin{eqnarray} \label{CS}
\Delta\cal{L}_{CS}&=&\frac{1}{4}\chi\emph{\~{F}}^{\mu\nu}(\nabla_{\nu}\emph{p}_{\mu}-\nabla_{\mu}\emph{p}_{\nu})
 \nonumber\\
 &=& \frac{1}{4}\chi\emph{\~{F}}^{\mu\nu}(\partial_{\nu}\emph{p}_{\mu}-\partial_{\mu}\emph{p}_{\nu})\,.
\end{eqnarray}
 From Eq.(\ref{current}), we have that
 $\partial_0 \emph{p}_i=0 $, $\partial_i
\emph{p}_0=\partial_i n_{\nu}(z)=0$ and $\partial_i \emph{p}_j=0$, which
guarantee $\Delta\cal{L}_{CS}$ in Eq. (\ref{CS}) is zero
due to the anti-symmetric property of $\emph{\~{F}}^{\mu\nu}$.
Consequently, we obtain that
$\emph{\~{F}}^{\mu\nu}(\partial_{\nu}\emph{p}_{\mu}-\partial_{\mu}\emph{p}_{\nu})=
\emph{\~{F}}^{\mu\nu}(\nabla_{\nu}\emph{p}_{\mu}-\nabla_{\mu}\emph{p}_{\nu})=0$
for the co-moving frame.
 We remark that  Eq. (\ref{Lcs}) is not formly gauge invariance.
 %always satisfied in general. 
 % The form in Eq. (\ref{Lcs}) is only true in comoving frame. 
 In general, to maintain the gauge invariance,
 we have to introduce the St$\ddot{u}$ckelberg field \cite{Geng:2007va,jackiw}.
%Therefore, 
%maintained \cite{added}. 
Moreover, the existence of a non-zero component $j_{\nu}^{0}$ in Eq. (\ref{Delta-n}) would violate Lorentz invariance \cite{r-3}.

It should be emphasized
%is interesting to point out
 that the
Chern-Simons like term in Eq. (\ref{Lcs}) is $P$ and $C$ odd but
$CPT$ even due to the $C$-odd vector current of $j_{\mu}$ in
Eq. (\ref{nu-current}), whereas the original one
in Ref. \cite{r-3} is $CPT$-odd \cite{coleman}.
It is clear that ${\cal L}_{CS}$ in  Eq. (\ref{Lcs}) is a
dimension-6 operator and it must be suppressed by two powers of
the mass scale $M$.

\section{Cosmological Birefringence}
Following
 Refs. \cite{r-4,r-3},
 the change  in the
position angle  of the polarization plane $\Delta\alpha$
 at redshift $z$
 due to our Chern-Simons-like term
  is given by
\begin{equation} \label{angle-1}
\Delta\alpha=\frac{1}{2}\frac{\beta}{M^2}\int \Delta n_{\nu}(t)
\frac{\textit{d}t}{R(t)}\,.
\end{equation}
To find out $\Delta\alpha$, we need to know the neutrino asymmetry
in our Universe, which is strongly constrained by the BBN
abundance of $^4$He. It is known that for a lepton flavor, the
asymmetry is given by: \cite{r-6,r-6r}
\begin{eqnarray} \label{asym-1}
\eta_{\ell}&=&\frac{n_{\ell}-n_{\bar{\ell}}}{n_{\gamma}}\;=\;\frac{1}{12\zeta(3)}
\left(\frac{T_{\ell}}{T_{\gamma}}\right)^3 (\pi^2
\xi_{\ell}+\xi_{\ell}^3)\,, \end{eqnarray} where $n_{i}\ (i=\ell,\gamma)$
are the $\ell$ flavor lepton and photon number densities, $T_{i}$
are the corresponding temperatures and
$\xi_{\ell}\equiv\mu_{\ell}/T_{\ell}$ is the degeneracy parameter.

As shown
%emphasized
by Serpico and Raffelt \cite{r-6}, the lepton
asymmetry in our Universe resides in neutrinos because of the
charge neutrality, while the neutrino number asymmetry depends
only on the electron-neutrino degeneracy parameter $\xi_{\nu_{e}}$
since neutrinos reach approximate chemical equilibrium before BBN
\cite{r-11}. From Eq. (\ref{asym-1}), the neutrino number
asymmetry for a lightest and relativistic, say, electron neutrino
is then given by
\cite{r-6,r-6r,r-6more}:
\begin{equation} \label{asym-2}
\eta_{\nu_e}\simeq 0.249 \xi_{\nu_{e}}
\end{equation}
where we have assumed $(T_{\nu_{e}}/T_{\gamma})^3=4/11$.
Note that the current
bound on the degeneracy parameter is $-0.046<\xi_{\nu_{e}}<0.072$
for a $2\sigma$ range of the baryon asymmetry
\cite{r-6,r-6r}. From Eqs. (\ref{Delta-n}),  (\ref{asym-1}) and (\ref{asym-2}),
we obtain
\begin{eqnarray} \label{asym-3}
\Delta n_{\nu}&
% \equiv & n_{\nu_e}-n_{\bar{\nu_e}} \\ \nonumber
         \simeq & 0.061\xi_{\nu_{e}}T_{\gamma}^3\,,
%                  n_{\gamma}
%          \nonumber\\
%         & = & 0.249 \xi_{\nu_{e}} \frac{\zeta(3)}{\pi^2} 2 T_{\gamma}^3
\end{eqnarray}
where we have used $n_{\gamma}=2\zeta(3)/\pi^2 \  T_{\gamma}^3$.
For a massless particle, after the decoupling, the evolution of its
temperature is given by
\cite{r-7}
\begin{eqnarray} \label{temp-1}
T R &=& T_D R_D\,,
\end{eqnarray}
where $T_D$ and $R_D$ are the temperature and scale factor at
 decoupling, respectively.
In particular, for $R=1$ at the present time, the photon temperature $T_{\gamma}^{\prime}$ of the red shift $z$  is
\begin{eqnarray}
 \label{photon}
T_{\gamma}&=&\frac{T_D R_D}{R}=T_{\gamma}^{\prime}(1+z)\,.
\end{eqnarray}
Then,  Eq. ({\ref{angle-1}) becomes
\begin{eqnarray}
 \label{angle-2}
\Delta\alpha &=&
%\frac{1}{2}\frac{\beta}{M^2}\int \Delta
%n_{\nu}(t) \frac{\textit{d}t}{R(t)} \nonumber\\ \nonumber
% &=& \frac{1}{2}\frac{\beta}{M^2} 0.249 \xi \frac{\zeta(3)}{\pi^2} 2
% T_{\gamma_{toady}}^3 \int_{t_*}^{t_{today}} (1+z)^3
% \frac{\textit{d}t}{R} \\ \nonumber
% &=& \frac{1}{2}\frac{\beta}{M^2} 0.249 \xi \frac{\zeta(3)}{\pi^2} 2
% T_{\gamma_{toady}}^3 \int_{R_*}^1 (1+z)^3
% \frac{\textit{d}R}{H(z)R^2}\\
% &=&
\frac{\beta}{M^2} 0.030 \xi_{\nu_{e}} (T_{\gamma}^{\prime})^3 \int_{0}^{z_*} (1+z)^3
 \frac{\textit{d}z}{H(z)}\,,
\end{eqnarray}
where 
%$H(z)$ is given by
%\begin{equation} \label{H}
$H(z)=H_0(1+z)^{3/2}$
%\end{equation}
in a flat and matter-dominated Universe and
 $H_0=2.1332 \times 10 ^{-42}h\ GeV$ is the Hubble constant
with  $h\simeq 0.7$ at the present.
%Then  Eq. ({\ref{angle-2}) is
%\begin{eqnarray} \label{angle-3}
%\Delta\alpha &=&\frac{1}{2}\frac{\beta}{M^2} 0.249 \xi
%\frac{\zeta(3)}{\pi^2} 2
% T_{\gamma_{toady}}^3 \int_{z_*}^0 (1+z)^3
% \frac{\textit{d}z}{H_0(1+z)^{3/2}} \\ \nonumber
% &=& \frac{1}{2}\frac{\beta}{M^2} 0.249 \xi
%\frac{\zeta(3)}{\pi^2} 2
% T_{\gamma_{toady}}^3 \frac{1}{H_0}\frac{2}{5}[(1+z_*)^{5/2}-1]
%\end{eqnarray}
We note that
as the rotation angle in Eq. (\ref{angle-2}) is mainly generated at the last scattering surface, there is no rotation of the large-scale CMB polarization which is generated by reionization at $z\sim 10$. However,
due to the accuracy level of current CMB polarization data, we
  have assumed a constant rotation angle over all angular scales.
 Finally, by taking $1+z_*=(1+z)_{decoupling}\simeq 1100$ at the photon
decoupling and $T_{\gamma}^{\prime}=2.73K$,
we get \begin{eqnarray} \label{angle-4} \Delta\alpha &\simeq& 4.2\times
10^{-2}\beta\left({\xi_{\nu_{e}}\over 0.001}\right)
\left({10\,TeV\over M}\right)^{2}\,. \end{eqnarray} As an illustration, for
example, by taking $\beta\sim 1$, $M\sim 10\ TeV$ and
$\xi_{\nu_{e}}\sim \pm 10^{-3}$, we get $\Delta \alpha
\sim\pm4\times 10^{-2}$, which could explain the results in Ref.
\cite{r-2}. We note that a sizable $\Delta \alpha$ could be
still conceivable even if the neutrino asymmetry is small. In that case,
the scale parameter $M$ has to be smaller.

\section{Summary}

%In summary, w
In Ref. \cite{Geng:2007va},
we have proposed a new type of effective interactions
in terms of the
$CPT$-even dimension-six Chern-Simons-like term,
which could originate from superstring theory,
 to generate
the cosmological birefringence. 
%We have pointed out that our effective interaction
%could originate from superstring theory, which may be viewed as a strong
%motivation of the present study.
To induce a sizable
rotation polarization angle in the CMB data,  a non-zero
neutrino number asymmetry is needed. 
%Finally, w
We remark that
the Planck Surveyor \cite{Planck} will reach a sensitivity of $\Delta \alpha$
at levels of $10^{-2}-10^{-3}$ \cite{WTNi-r,Lue}, while a dedicated future experiment on
the cosmic microwave background radiation polarization would reach
$10^{-5}-10^{-6}$ $\Delta \alpha$-sensitivity \cite{WTNi-r}.\\

%\newpage
\noindent
{\bf Acknowledgments}
%\\

Two of us (CQG and SHH) would like to thank the organizers: Prof. Manu Paranjape and
Prof. Richard MacKenzie for 
the hospitality during the conference. 
%We thank Prof. W.T. Ni, Dr. T.C. Yuan, Dr. Y.K. Hsiao 
%and Dr. W. Liao for useful discussions.
This work is supported in part by
the Natural Science and Engineering Council of Canada
and
the National Science Council of
R.O.C. under Contract \#s: NSC-95-2112-M-007-059-MY3.
%and NSC-96-2918-I-007-010.

\end{document}